\documentclass[aps,prb,amsmath,amssymb,reprint]{revtex4-1}
\usepackage{graphicx}
\usepackage{color}
\usepackage{bm}
\usepackage{mathtools}
\usepackage{hyperref}

\makeatother
\begin{document}

\title{Spontaneous symmetry breaking without ground state degeneracy in generalized $N$-state clock model}
\author{Yaozong Hu}
\author{Haruki Watanabe}
\altaffiliation{Department of Applied Physics, The University of Tokyo.}

\date{\today}

\begin{abstract}
Spontaneous symmetry breaking is ubiquitous phenomenon in nature.
One of the defining features of symmetry broken phases is that the large system size limit and the vanishing external field limit do not commute.
In this work, we study a family of extensions of the $N$-state clock model.
We find that the exact symmetry and the ground state degeneracy under the periodic boundary condition heavily depend on the system size, although the model has the manifest translation symmetry. 
In particular, the ground state can be unique and all excitations are gapped even when the phase exhibits non-commutativity of the two limits.
Our model hence poses a question on the standard understanding of spontaneous symmetry breaking.
\end{abstract}

\maketitle

\section{Introduction}
Spontaneous breaking of symmetry is a phenomenon in which the symmetry of the Hamiltonian or the Lagrangian of the system is not respected by physical states.
It underlies many phases of matter such as crystals, magnets, and superfluids, and has been studied for a long time.

\begin{figure}[t]
\includegraphics[width=\columnwidth]{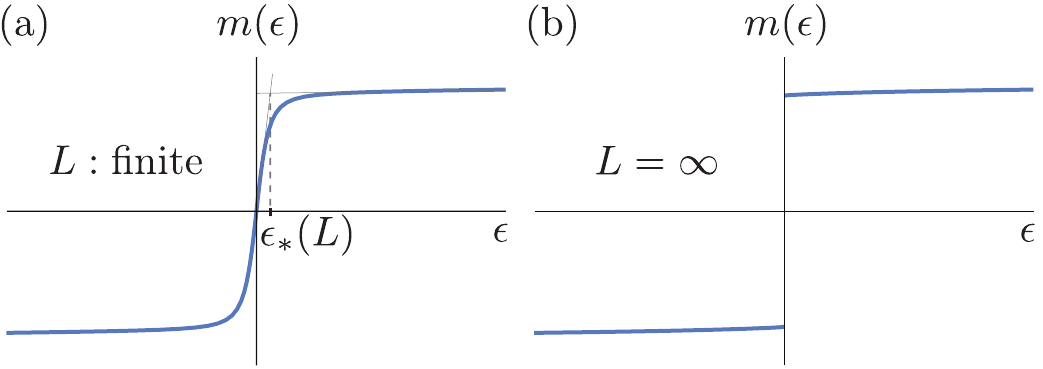}
\caption{\label{fig0} 
Illustration of the typical behavior of an order parameter $m(\epsilon)$ as a function of symmetry breaking field $\epsilon$ in an ordered phase. Panel (a) is for a finite $L$ and the curve is continuous, while (b) is for the large $L$ limit and the curve is discontinuous at $\epsilon=0$. In the panel (a), $\epsilon_*(L)$ represents the characteristic value of $\epsilon$ that separates the linear-response regime [$m(\epsilon)\propto\epsilon$] and the saturation regime. The discontinuity in (b) can be rephrased as $\lim_{L\rightarrow\infty}\epsilon_*(L)=0$.}
\end{figure}

To review its basic understanding, let us consider a quantum system at the zero temperature $T=0$.
For simplicity here we consider a \emph{discrete} symmetry group, not a continuous one.
Suppose that the symmetry of the system is spontaneously broken down to its subgroup.
Empirically, such a phase commonly exhibits the following features:
(i) The $M$ lowest energy eigenstates in a finite system, which respect all of the original symmetries, are given as superpositions of $M$ symmetry-breaking states.
Here, $M$ represents the number of broken symmetry elements.
(ii) The splitting of the $M$ lowest energy eigenvalues are exponentially suppressed with the system size $V$, while the excitation gap to the next energy level is $O(1)$.
(iii) When a symmetry-breaking field $\epsilon$ that favors one of the symmetry breaking states is introduced, the large system size limit ($V\rightarrow\infty$) and the vanishing field limit ($\epsilon\rightarrow+0$) do not commute, as illustrated in Fig.~\ref{fig0}.
The first two properties constitute the $M$-fold ground state degeneracy in the symmetry broken phase that is protected by the broken symmetry of the system. The last feature implies the instability of the symmetric ground state toward an ordered state, which explains why cat states [i.e., the symmetric superpositions described in (i)] are fragile and never be realized in nature. Since the property (iii) is sometimes taken as the definition of spontaneous-symmetry breaking~\cite{auerbach1998interacting,altland_simons_2010,TasakiBook}, one may expect that properties (i) and (ii) follow automatically as consequences of (iii). 

The transverse-field Ising model is a prototypical quantum spin model that exhibits spontaneous breaking of $\mathbb{Z}_2$ symmetry and quantum phase transition to a symmetry unbroken phase~\cite{SachdevBook}.  Its generalization to $N$-level spin system with $\mathbb{Z}_N$ symmetry is called the $N$-state clock model.  The  $N$-state clock models show all the three features associated with spontaneous symmetry breaking summarized above, as we review in Sec.~\ref{reviewa1} below.
The quantum phase transition in the $N$-state clock model belongs to the same universality class as in a recent experimental study~\cite{Bernien2017} of the cold atom system~\cite{PhysRevB.98.205118}.

In this work, we introduce a generalization of the $N$-state clock model that show several intriguing behaviors.
This model is hinted by a recent study~\cite{2211.00299} of generalized $\mathbb{Z}_N$ toric code~\cite{doi:10.1063/1.1499754,KITAEV20032}, whose ground state was shown to be unique for a sequence of system size despite its topological order.  
Our model consists of at most two-spin interactions and the nearest-neighbour interaction contains an integer parameter $a=1,2,\cdots,N$. The standard $N$-state clock model corresponds to the $a=1$ case. When
$a\neq1$, the exact symmetry and ground state degeneracy heavily depend on the system size under the periodic boundary condition, although the model has the translation symmetry. In particular, even when a symmetry breaking occurs in the sense that the limits of large system size and vanishing symmetry-breaking field do not commute, the ground state can be unique and excitations can be gapped, depending on the system size.  Hence, this model can be regarded as an example that exhibits the feature (iii) without (i) and (ii).

Some previous studies of the transverse-field Ising model observed similar behaviors, but these models are different from ours in an essential way.
For example, in the antiferromagnetic Ising model, the ground states are two-fold degenerate and excitations are gapped for even $L$ and excitations are gapless for odd $L$~\cite{Dong_2016}. However, it cannot realize a unique ground state with an excitation gap. 
In contrast, if a symmetry-breaking external field is applied to the two spins at the ends of an open ferromagnetic Ising chain, the ground state can be unique and excitations are gapped even in the ordered phase~\cite{PhysRevE.98.032124}. 
In fact, as we will see below, our model can, in some cases, be mapped to the standard $N$-state clock model with a twisted boundary condition, which may be understood as an $N$-level version of the Ising chain with symmetry-breaking boundary condition. However, such a model lacks the translation symmetry unlike our model.

\section{Generalized $N$-state clock model}
In this section, we present the definition of our generalized $N$-state clock model and examine its symmetries.
\subsection{Definitions}
We consider one dimensional system consisting of $L$ spins.
$N$-level ($N\geq2$) spin operators are generalizations of $S=1/2$ spin operators, satisfying
\begin{align}
&\hat{Z}_i\hat{X}_{i'}=\omega^{\delta_{i,i'}}\hat{X}_{i'}\hat{Z}_i,\quad \omega\coloneqq e^{2\pi i/N},\label{spinalgebra}
\end{align}
and $\hat{Z}_i^N=\hat{X}_i^N=1$ for $i,i'=0,1,2,\cdots,L-1$. The basis states $\{|\omega^\ell\rangle_i\}_{\ell=0}^{N-1}$ for the $i$-th spin are defined by  $\hat{Z}_i|\omega^\ell\rangle_i=\omega^\ell|\omega^\ell\rangle_i$ and $\hat{X}_i|\omega^\ell\rangle_i=|\omega^{\ell+1}\rangle_i$.  The total Hilbert space dimension is $N^{L}$. 

The Hamiltonian of our model reads as
\begin{align}
\hat{H}&\coloneqq-\frac{1}{2}\sum_{i=0}^{L-1}\left[(\hat{Z}_i^{-a}\hat{Z}_{i+1}+\text{h.c.})+g(\hat{X}_i+\text{h.c.})\right],\label{generalizedmodel}
\end{align}
where $a=1,2,\cdots,N$ is an integer parameter that specifies the nearest-neighbor interaction. 
The standard $N$-state clock model corresponds to $a=1$.
The transverse field $g$ is assumed to be nonnegative.
The periodic boundary condition is imposed so that $\hat{Z}_{i+L}=\hat{Z}_i$ and $\hat{X}_{i+L}=\hat{X}_i$.  The energy eigenstate of $\hat{H}$ is written as $|\Phi_n\rangle$ ($n=0,1,2,\cdots$) in the increasing order of the energy eigenvalues $E_0\leq E_1\leq E_2\cdots$. 

When $a=N$, the first term becomes the longitudinal field term $\sum_{i=1}^N(\hat{Z}_i+\text{h.c.})$ and the model is trivial. When $2\leq a\leq N-1$, the properties of this model generally exhibit a nontrivial dependence on the system size, as we shall see below. 

In our numerical study, exact diagonalization is performed up to $L=23$ for $N=2$, $L=15$ for $N=3$, and $L=12$ for $N=4$.
Larger system sizes for $N=3$ are handled by the density matrix renormalization group (DMRG) method using ITensor~\cite{itensor}.

\subsection{Symmetries}
The symmetries of the generalized model can be divided into two classes: those which always exist, and those which might be absent depending on the system size $L$ and the parameter $a$. For example, the model always has the translation symmetry $\hat{T}$, defined by
\begin{align}
\hat{T}\hat{Z}_i\hat{T}^{-1}=\hat{Z}_{i+1},\quad \hat{T}\hat{X}_i\hat{T}^{-1}=\hat{X}_{i+1}.\quad\hat{T}^L=1.\label{deftranslation}
\end{align}
The model also has the charge flip symmetry $\hat{C}$~\cite{ORTIZ2012780,PhysRevB.95.235127} and the time-reversal symmetry $\hat{K}$~\cite{PhysRevB.98.205118}, defined by
\begin{align}
&\hat{C}\hat{Z}_i\hat{C}^{-1}=\hat{Z}_i^{-1},\quad\hat{C}\hat{X}_i\hat{C}^{-1}=\hat{X}_i^{-1},\quad\hat{C}^2=1\label{defC},\\
&\hat{K}\hat{Z}_i\hat{K}^{-1}=\hat{Z}_i^{-1},\quad\hat{K}\hat{X}_i\hat{K}^{-1}=\hat{X}_i,\quad\hat{K}^2=1\label{defK}.
\end{align}
To be consistent with the spin algebra in Eq.~\eqref{spinalgebra}, $\hat{C}$ is unitary and $\hat{K}$ is anti-unitary. These symmetries all commute with each other.

Furthermore, depending on $L$ and $a$, the model has a discrete spin-rotation symmetry generated by
\begin{align}
\hat{X}\coloneqq\prod_{i=0}^{L-1} \hat{X}_i^{a^i}.\label{defX2}
\end{align}
We find $\hat{X}^N=1$ and 
\begin{align}
\hat{X}^n\hat{H}\hat{X}^{-n}-\hat{H}=\frac{1-\omega^{n(a^L-1)}}{2}\hat{Z}_{L-1}^{-a}\hat{Z}_{0}+\text{h.c.}
\end{align}
Therefore, if $n$ is set to be
\begin{align}
n\coloneqq \frac{N}{\mathrm{gcd}(a^L-1,N)},\label{defn}
\end{align}
then $n(a^L-1)=0\mod N$ and $[\hat{X}^n,\hat{H}]=0$. Here, $\mathrm{gcd}(p,q)$ represents the greatest common divisor of integers $p$ and $q$. Therefore, given $N$, $a$, and $L$, the exact spin-rotation symmetry of the model is given by
\begin{align}
\mathbb{Z}_{\mathrm{gcd}(a^L-1,N)},\label{general1}
\end{align}
implying that the ground state degeneracy in the ferromagnetically ordered phase is 
\begin{align}
N_{\mathrm{deg}}=\mathrm{gcd}(a^L-1,N).\label{general2}
\end{align}

The operator $\hat{X}$ satisfies
\begin{align}
&\hat{C}\hat{X}\hat{C}^{-1}=\hat{X}^{-1},\quad\hat{K}\hat{X}\hat{K}^{-1}=\hat{X},\\
&\hat{T}^{-1}\hat{X}\hat{T}= \hat{X}_{L-1}\prod_{i=1}^{L-1} \hat{X}_{i-1}^{a^i}=\hat{X}^{a}\hat{X}_{L-1}^{1-a^{L}}.\label{TXT2}
\end{align}
From the second relation, it follows that $[\hat{X}^n,\hat{T}]\neq0$, implying that the spin-rotation symmetry $\hat{X}^n$ is not a genuine internal symmetry unless $a=1$.

When $a=1$, $N-1$, or $N$, the model also has the spatial inversion symmetry 
\begin{align}
&\hat{I}\hat{Z}_i\hat{I}^{-1}=\hat{Z}_{-i},\quad \hat{I}\hat{X}_i\hat{I}^{-1}=\hat{X}_{-i},\quad \hat{I}^2=1.\label{inversion}
\end{align}
which is explicitly broken for $a=2,3,\cdots,N-2$.

\subsection{Duality}
As is well-known (see e.g. Ref.~\onlinecite{PhysRevB.98.205118}), the standard $N$-state clock model has a duality between $g$ and $1/g$, which persists in the generalized model with $a\neq1$ as we shall see now. Dual spin operators are defined by the nonlocal transformation
\begin{align}
\hat{\tilde{Z}}_i\coloneqq\prod_{i'=i}^{L-1}\hat{X}_{i'}^{a^{i'-i}},\quad \hat{\tilde{X}}_i\coloneqq\hat{Z}_{i-1}^{a}\hat{Z}_{i}^{-1},\label{dual}
\end{align}
which satisfy
\begin{align}
&\hat{\tilde{Z}}_i^N=\hat{\tilde{X}}_i^N=1,\quad\hat{\tilde{Z}}_i\hat{\tilde{X}}_{i'}=\omega^{\delta_{i,i'}}\hat{\tilde{X}}_{i'}\hat{\tilde{Z}}_i,\label{ns2}\\
&\hat{\tilde{Z}}_{i'}\hat{\tilde{X}}_{0}=\omega^*\hat{\tilde{X}}_{0}\hat{\tilde{Z}}_{i'},\quad\hat{\tilde{Z}}_{0}\hat{\tilde{X}}_{0}=\hat{\tilde{X}}_{0}\hat{\tilde{Z}}_{0}\label{ns4}
\end{align}
for $i=0,1,2,\cdots,L$ and $i'=1,2,\cdots,L-1$. This map converts $\hat{H}$ to 
\begin{align}
\hat{H}&=-\frac{g}{2}\sum_{i=0}^{L-2}\left[(\hat{\tilde{Z}}_i\hat{\tilde{Z}}_{i+1}^{-a}+\text{h.c.})+(1/g)(\hat{\tilde{X}}_i+\text{h.c.})\right]\notag\\
&\quad\quad-\frac{1}{2}\left[(\hat{\tilde{X}}_{L-1}+\text{h.c.})+g(\hat{\tilde{Z}}_{L-1}+\text{h.c.})\right].\label{DualH}
\end{align}
This expression coincides with $g\hat{H}(1/g)$ except for the boundary terms and the spatial inversion.

\begin{figure}[t]
\includegraphics[width=\columnwidth]{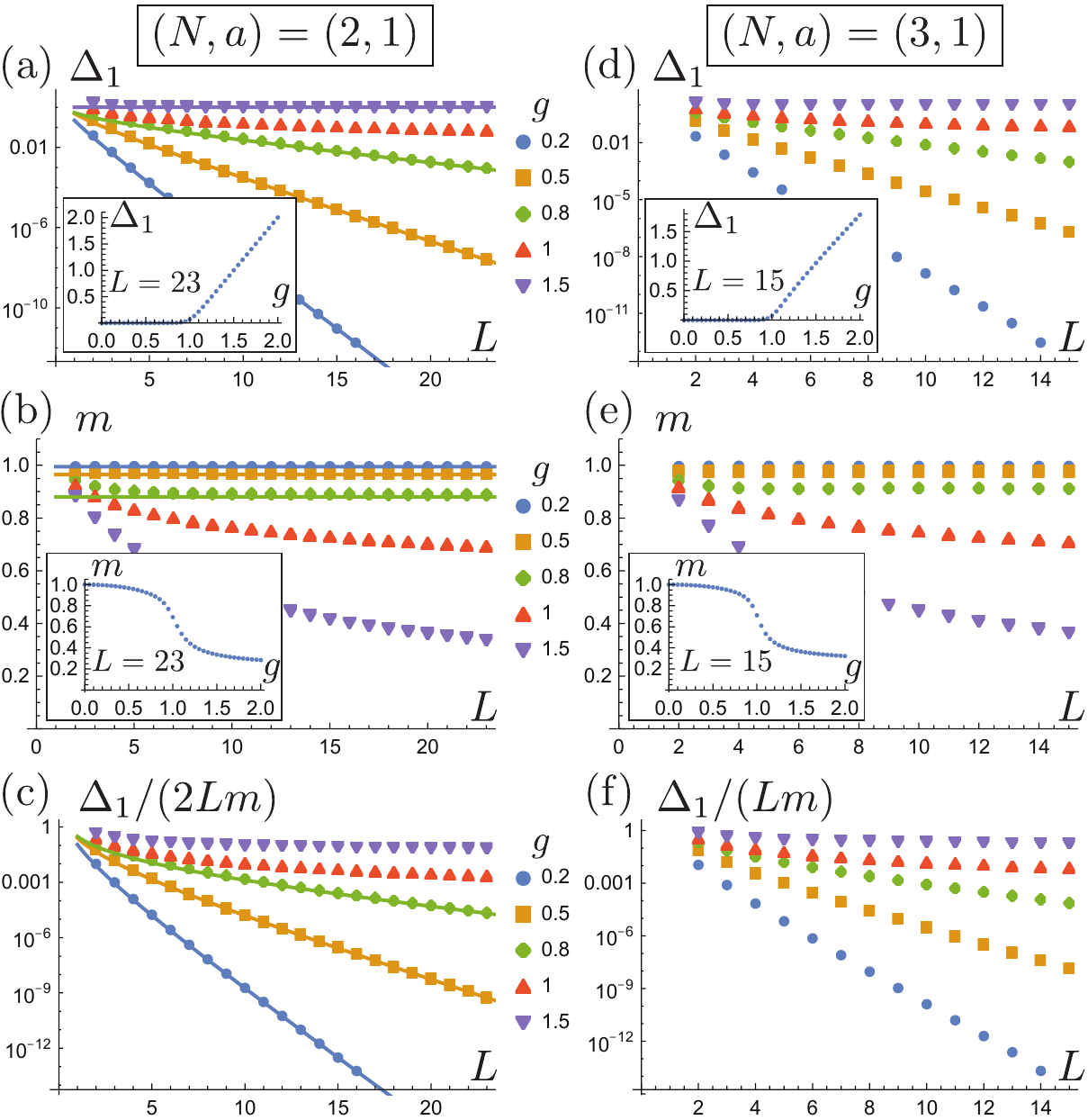}
\caption{\label{fig1} 
Exact-diagonalization results for the standard $N$-state clock model with $N=2$ [(a)--(c)] and $N=3$ [(d)--(f)].
(a),(d): The energy difference $\Delta_1\coloneqq E_1-E_0$ between the ground state and the first excited state in a finite system. 
(b),(e): The long-range order $m\coloneqq \sqrt{\langle \Phi_0|\hat{z}^\dagger\hat{z}|\Phi_0\rangle}$. 
(c),(f): $\Delta_1/(2Lm)$ [(c)] and $\Delta_1/(Lm)$ [(f)] that approximate $\epsilon_*(L)$.
The insets in (a),(b),(d),(e) show the $g$ dependence.
The curves in (a)-(c) are the analytic expressions in Eqs.~\eqref{mgN2},\eqref{Delta1N2}, and \eqref{epsilonstarN2}.
}
\end{figure}

\section{Examples}
In this section we study the properties of the generalized models for several representative values of $a$.
\subsection{$a=1$} 
\label{reviewa1}
Let us begin by reviewing the standard $N$-state clock model. When $a=1$, $\hat{X}$ in Eq.~\eqref{defX2} has no position dependence and commutes with $\hat{H}$ regardless of $L$, generating a $\mathbb{Z}_N$ symmetry [i.e., $n=1$ in Eq.~\eqref{defn}].
When $1\gg g\geq0$, the $\mathbb{Z}_N$ symmetry is spontaneously broken, while no symmetries are broken when $g\gg 1$. A continuous phase transition occurs at $g=1$ for $N=2,3,4$~\cite{ORTIZ2012780,PhysRevB.100.094428}. There are two BKT transitions at $g=g_1$ ($1>g_1\geq0$) and $1/g_1$ for $N\geq5$, as suggested by the aforementioned duality~\cite{ORTIZ2012780,PhysRevB.100.094428}.

\subsubsection{Order parameter, long-range order, and finite size splitting}
To diagnose spontaneous breaking of $\mathbb{Z}_N$ symmetry, let us introduce an order parameter
\begin{align}
\hat{z}\coloneqq \frac{1}{L}\sum_{i=0}^{L-1}\hat{Z}_i,
\end{align}
which transforms linearly under $\hat{X}$: 
\begin{align}
\hat{X}^\dagger\hat{z}\hat{X}=\omega\hat{z}.\label{XzX}
\end{align}
When $g=0$, product states 
\begin{align}
|\phi_\ell\rangle\coloneqq\bigotimes_{i=0}^{L-1}|\omega^\ell\rangle_i=\hat{X}^\ell|\phi_0\rangle
\end{align}
 ($\ell=0,1,2,\cdots,N-1$) are symmetry-breaking ground states, characterized by the expectation value $\langle\phi_\ell|\hat{z}|\phi_\ell\rangle=\omega^\ell$. The $N$-fold degeneracy is guaranteed by the $\mathbb{Z}_N$ symmetry: $[\hat{H},\hat{X}]=0$.
 In addition to the time-reversal symmetry $\hat{K}$ and the translation symmetry $\hat{T}$, the $\mathbb{Z}_2$ symmetry generated by $\hat{C}_\ell\coloneqq\hat{X}^\ell\hat{C}\hat{X}^{-\ell}$ remains unbroken for each $\ell=0,1,2,\cdots,N-1$. The gap to the $N+1$-th state is given by $2(1-\cos\frac{2\pi}{N})$.

When $g\neq 0$ but still in the range $1\gg g>0$,  the $N$ lowest energy eigenstates remain separated by other excited states by an $O(1)$ excitation gap. In particular, the ground state $|\Phi_0\rangle$ in a finite system can be approximated by the linear combination $N^{-1/2}\sum_{\ell=0}^{N-1}|\phi_\ell\rangle+O(g)$. This state is symmetric, $\hat{X}|\Phi_0\rangle=|\Phi_0\rangle$, and the expectation value of the order parameter vanishes, $\langle\Phi_0|\hat{z}|\Phi_0\rangle=0$. Instead, this state has a long-range correlation, which can be captured by the large $L$ limit of 
\begin{align}
m&\coloneqq\sqrt{\langle \Phi_0|\hat{z}^\dagger\hat{z}|\Phi_0\rangle}.
\label{defm}
\end{align}
For example, when $N=2$, an analytic expression is known\cite{PFEUTY197079}
\begin{align}
\lim_{L\rightarrow\infty}m=(1-g^2)^{1/8}.\label{mgN2}
\end{align}
A nonzero value of the long-range order $m$ in the large system size limit implies spontaneous breaking of the $\mathbb{Z}_N$ symmetry~\cite{doi:10.1143/JPSJ.58.3894}.

The finite-size splitting of energy eigenvalues of lowest $N$ eigenstates is typically the order of $g^L=e^{-L/\xi}$ ($\xi\coloneqq-1/\log g$), which is exponentially suppressed with the system size. This can be most easily understood by the perturbation theory from the $g=0$ point, since at least $L$-th order perturbation is needed to generate nonzero matrix elements among $|\phi_\ell\rangle$ ($\ell=0,1,2,\cdots,N-1$). For example, when $N=2$, the asymptotic behavior for a large $L$ is given by\cite{PhysRevB.35.7062}
\begin{align}
\Delta_1\coloneqq E_1-E_0\simeq2\sqrt{\frac{1-g^2}{\pi L}}g^L \big[1+O(L^{-1})\big].\label{Delta1N2}
\end{align}
For reader's convenience, we include the derivation of Eqs.~\eqref{mgN2} and \eqref{Delta1N2} in Appendix~\ref{appJW}. By exact diagonalization, we confirm the validity of these analytic expressions  by numerics in Fig.~\ref{fig1} (a)--(c).
For $N\geq 3$, such expressions are not known but we numerically demonstrate that the $N=3$ case behaves similarly in Fig.~\ref{fig1} (d)--(f).

\subsubsection{Symmetry breaking field}
\label{secSSF1}
Another way to detect spontaneous symmetry breaking is to apply a symmetry-breaking field $\epsilon\geq0$~\cite{TasakiBook}. We replace the Hamiltonian $\hat{H}$ with
\begin{align}
\hat{H}^{(\ell_0)}(\epsilon)&\coloneqq\hat{H}-\frac{1}{2}\epsilon L\left(\omega^{-\ell_0}\hat{z}+\text{h.c.}\right).\label{He}
\end{align}
The parameter $\ell_0=0,1,\cdots,N-1$ selects the symmetry breaking state favored by $\epsilon>0$. 
As far as the $\mathbb{Z}_N$ symmetry generated by $\hat{X}$ is exact, all values of $\ell_0$ are equivalent in the sense that they are related by the $\mathbb{Z}_N$ symmetry.

\begin{figure}[t]
\includegraphics[width=\columnwidth]{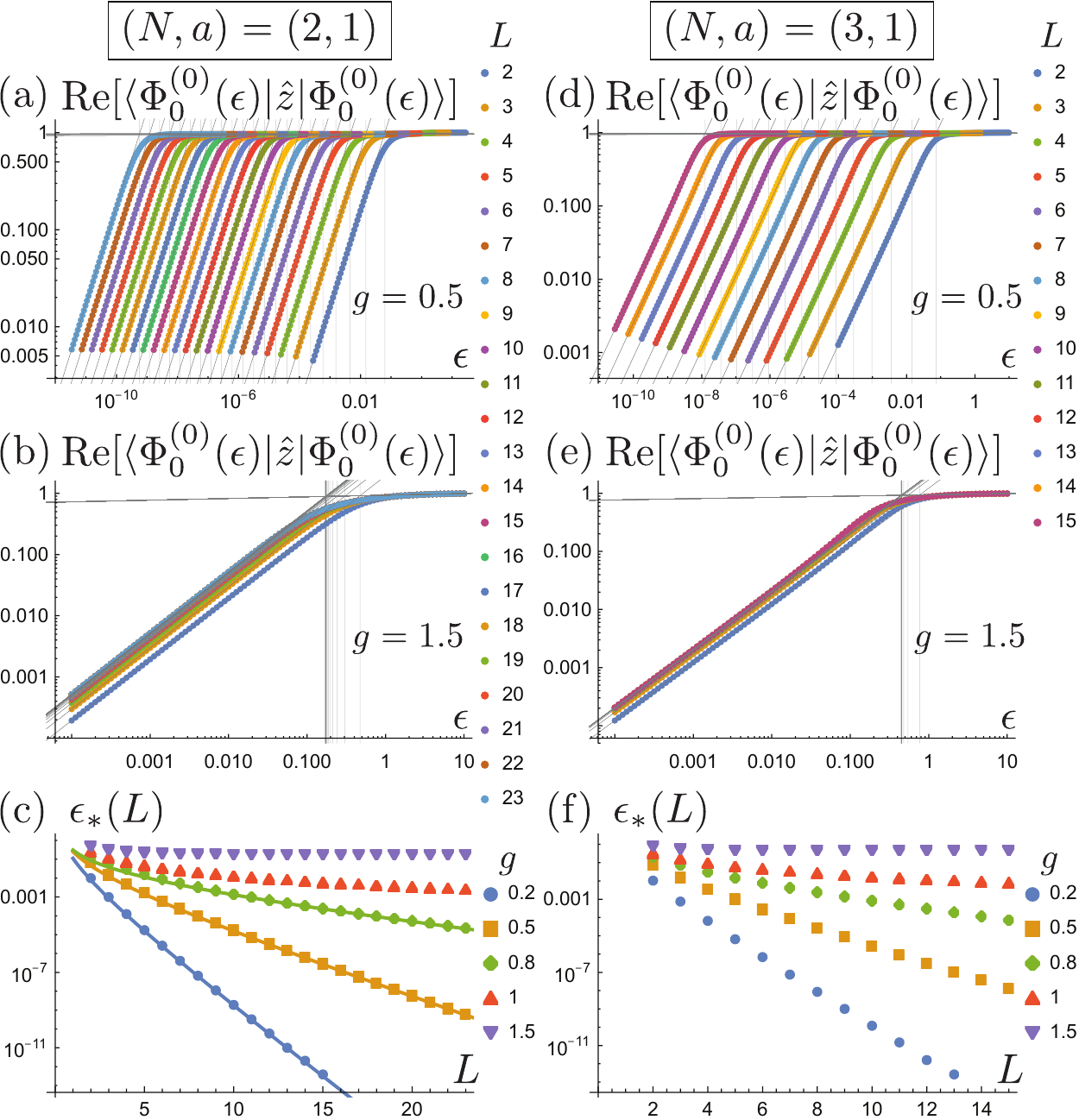}
\caption{\label{fig2} 
Exact-diagonalization results for the standard $N$-state clock model with symmetry-breaking field $\epsilon$ for $N=2$ [(a)--(c)] and for $N=3$ [(d)--(f)]. Here we set $\ell_0=0$ as an example. 
(a),(b),(d),(e): The order parameter $\text{Re}[\langle\Phi_0^{(0)}(\epsilon)|\hat{z}|\Phi_0^{(0)}(\epsilon)\rangle]$ for $g=0.5$ [(a),(c)] and $g=1.5$ [(b),(e)].
(c),(f): The magnetic field $\epsilon_*(L)$ at the transition point, which is determined by the crossing point of two fitting lines (gray lines) in the log-log plot of $\text{Re}[\langle\Phi_0^{(0)}(\epsilon)|\hat{z}|\Phi_0^{(0)}(\epsilon)\rangle]$.
The curves in (c) are the analytic expression in Eq.~\eqref{epsilonstarN2}.
}
\end{figure}

The effect of symmetry-breaking field can be understood analytically based on the effective Hamiltonian that focuses on the $N$ low-energy states. For example, for $N=2,3,4$, we find
\begin{align}
H_{\mathrm{eff}}^{(\ell_0)}(\epsilon)=&-\frac{c_N\Delta_1}{2}(X+\text{h.c.})-\frac{\epsilon Lm}{2}(\omega^{-\ell_0} Z+\text{h.c.})
\end{align}
in the basis of symmetry breaking states, where $c_2=1/2$, $c_3=2/3$, and $c_4=1$, and 
\begin{align}
X\coloneqq\begin{pmatrix}
&&&1\\
1&&&\\
&\ddots&&\\
&&1&\\
\end{pmatrix},
\quad Z\coloneqq\begin{pmatrix}
1&&&\\
&\omega&&\\
&&\ddots&\\
&&&\omega^{N-1}
\end{pmatrix}.
\end{align}
The first term describes the mixing due to $g\neq0$, and the second term favors the symmetry breaking state that matches the applied field.

For a small $\epsilon$, the order parameter $\text{Re}[\langle \omega^{-\ell_0} Z \rangle]$ exhibits the linear response $\text{Re}[\langle \omega^{-\ell_0} Z \rangle]\propto Lm^2\epsilon/\Delta_1$, 
while it is saturated $\text{Re}[\langle \omega^{-\ell_0} Z \rangle]\simeq m$ for a large $\epsilon$.
The transition occurs at $\epsilon=\epsilon_*(L)$ where the first term and second term balance. We find
\begin{align}
&\epsilon_*^{N=2}(L)\simeq\frac{\Delta_1}{2Lm}\simeq\frac{(1-g^2)^{3/8}}{\sqrt{\pi} L^{3/2}}g^L,\label{epsilonstarN2}\\
&\epsilon_*^{N=3,4}(L)\simeq\frac{\Delta_1}{Lm}.
\end{align}
It follows that
\begin{align}
\lim_{L\rightarrow\infty}\epsilon_*(L)=0,\label{noncommutative0}
\end{align}
implying the discontinuity in the expectation value of the order parameter as a function of $\epsilon$ in the thermodynamic limit. This observation suggests that the small $\epsilon$ limit and the large $L$ limit do not commute:
\begin{align}
&\lim_{\epsilon\rightarrow+0}
\lim_{L\rightarrow\infty}
\langle\Phi_0^{(\ell_0)}(\epsilon)|\hat{z}|\Phi_0^{(\ell_0)}(\epsilon)\rangle\neq0,\label{noncommutative1}\\
&\lim_{L\rightarrow\infty}
\lim_{\epsilon\rightarrow+0}
\langle\Phi_0^{(\ell_0)}(\epsilon)|\hat{z}|\Phi_0^{(\ell_0)}(\epsilon)\rangle=0,\label{noncommutative2}
\end{align}
where $|\Phi_0^{(\ell_0)}(\epsilon)\rangle$ is the unique ground state of $\hat{H}^{(\ell_0)}(\epsilon)$. 

We confirm this understanding by numerical calculations. The panels (a) and (d) in Fig.~\ref{fig2} for $N=2$ and $3$, respectively, demonstrate that the expectation value of the order parameter develops as the symmetry breaking field $\epsilon$ increases, and saturates around $\epsilon=\epsilon_*(L)$. The saturation field $\epsilon_*(L)$ gets smaller and smaller as the system size increases as shown in Fig,~\ref{fig2} (d),(f), and 
in the limit of the large system size. This is the numerical demonstration of the noncommutative nature of the two limits in Eqs.~\eqref{noncommutative1} and \eqref{noncommutative2}.

\subsection{$N$ is odd and $a=N-1$}
Next, let us study the simplest nontrivial case. When $a=N-1$, the Hamiltonian becomes
\begin{align}
\hat{H}&\coloneqq-\frac{1}{2}\sum_{i=0}^{L-1}\left[(\hat{Z}_i\hat{Z}_{i+1}+\text{h.c.})+g(\hat{X}_i+\text{h.c.})\right],\label{generalizedmodel2}
\end{align}
which is still manifestly translation invariant.

\subsubsection{$L$: even}
Let us first assume that $L$ is even. In this case, the model has a modified $\mathbb{Z}_N$ symmetry generated by
\begin{align}
\hat{X}\coloneqq\prod_{i=0}^{L-1} \hat{X}_i^{(-1)^i}=\hat{X}_0\hat{X}_1^\dagger \hat{X}_2\hat{X}_3^\dagger\cdots\hat{X}_{L-2}\hat{X}_{L-1}^\dagger.\label{defX3}
\end{align}
In other words, $n$ in Eq.~\eqref{defn} is $1$. The corresponding order parameter 
\begin{align}
\hat{z}&\coloneqq \frac{1}{L}\sum_{i=0}^{L-1}\hat{Z}_i^{(-1)^i}\notag\\
&=\frac{1}{L}(\hat{Z}_0+\hat{Z}_1^\dagger+\hat{Z}_2+\hat{Z}_3^\dagger+\dots+\hat{Z}_{L-2}+\hat{Z}_{L-1}^\dagger)\label{defOP2}
\end{align}
satisfies Eq.~\eqref{XzX}. This model can be mapped to the standard one with $a=+1$ by the $\hat{C}$ transformation in Eq.~\eqref{defC} for spins only on even sites. Therefore, as far as thermodynamic properties are concerned, the $a=N-1$ model should be equivalent to the standard $a=+1$ model.  In particular, when $N=3$, the $\mathbb{Z}_N$ symmetry of the model is spontaneously broken when $1>g\geq0$ and a phase transition to the paramagnetic phase occurs at $g=1$.

When $g=0$, the ferromagnetic state $|\phi_0\rangle\coloneqq\bigotimes_{i=0}^{L-1}|1\rangle_i$ is a ground state. $N$ distinct ground states can be generated as
\begin{align}
|\phi_\ell\rangle\coloneqq\hat{X}^\ell|\phi_0\rangle=\bigotimes_{i=0}^{L-1}|\omega^{\ell (-1)^i}\rangle_i,\label{gsr}
\end{align}
whose expectation value of order parameter is $\langle\phi_\ell|\hat{z}|\phi_\ell\rangle=\omega^\ell$ ($\ell=0,1,2,\cdots,N-1$).  Interestingly, $|\phi_{\ell}\rangle$ with $\ell\neq0$ is not translation invariant, i.e., $\hat{T}|\phi_{\ell}\rangle=|\phi_{-\ell}\rangle$. It is instead symmetric under a modified translation symmetry $\hat{T}_\ell\coloneqq\hat{T}\hat{X}^{-2\ell}$.

\begin{figure}[t]
\includegraphics[width=\columnwidth]{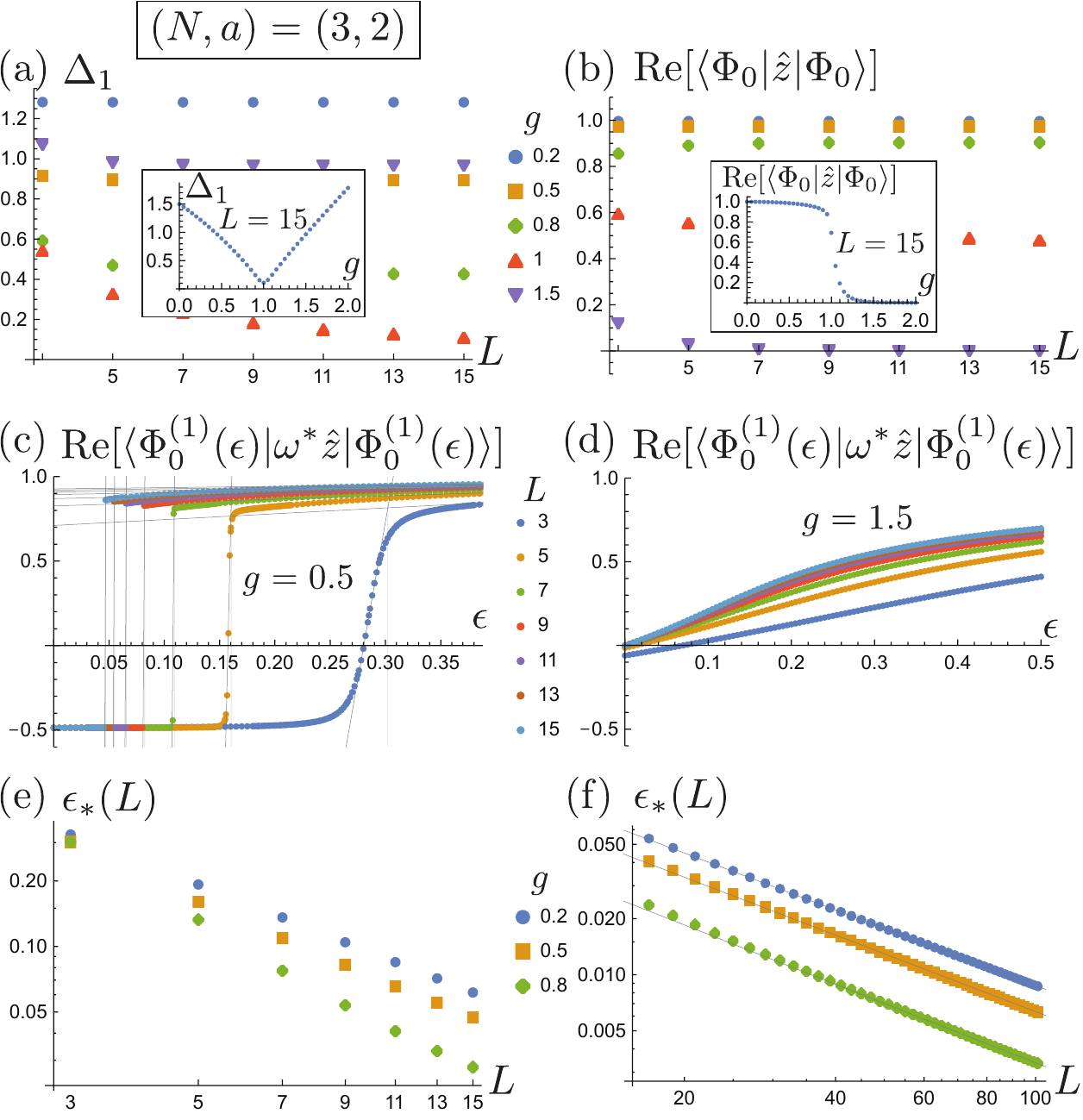}
\caption{\label{fig3} 
Exact diagonalization results [(a)--(e)] and DMRG results [(f)] for the $(N,a)=(3,2)$ case.
(a): The energy difference $\Delta_1=E_1-E_0$. 
(b): The order parameter $\text{Re}[\langle\Phi_0|\hat{z}|\Phi_0\rangle]$.
The insets in (a),(b) show the $g$ dependence.
(c),(d): $\text{Re}[\langle\Phi_0^{(1)}(\epsilon)|\omega^*\hat{z}|\Phi_0^{(1)}(\epsilon)\rangle]$ for $g=0.5$ [(c)] and $g=1.5$ [(d)].
(e): The magnetic field $\epsilon_*(L)$ at the transition point, which is determined by the crossing points of two fitting lines in panels (c).
(f): The same as (e) but for larger system size ($17\leq L\leq 101$) computed by DMRG. The fitting lines have slope $-1$ with few $\%$ error, confirming the $L^{-1}$ dependence of $\epsilon_*(L)$.
}
\end{figure}

\subsubsection{$L$: odd}
Next let us assume that $L$ is odd. In this case, 
\begin{align}
\hat{X}\coloneqq\prod_{i=0}^{L-1} \hat{X}_i^{(-1)^i}=\hat{X}_0\hat{X}_1^\dagger \hat{X}_2\hat{X}_3^\dagger\cdots\hat{X}_{L-2}^\dagger\hat{X}_{L-1}\label{defX4}
\end{align}
does not commute with $\hat{H}$ in Eq.~\eqref{generalizedmodel2}:
\begin{align}
\hat{X}^\dagger\hat{H}\hat{X}-\hat{H}=\frac{1-\omega^{2}}{2}\hat{Z}_{L-1}\hat{Z}_{0}+\text{h.c.}
\end{align}
Furthermore, $n$ in Eq.~\eqref{defn} is $N$ and $\hat{X}^n$ becomes the identity operator. As a consequence, the ground state is unique and excitations are gapped even in the range $1> g\geq0$. We show our numerical results for $N=3$ in Fig.~\ref{fig3} (a).

Unlike the even $L$ case, the unique ground state has a nonzero expectation value of the order parameter $\text{Re}[\langle\Phi_0|\hat{z}|\Phi_0\rangle]$ as shown in Fig.~\ref{fig3} (b), where
\begin{align}
\hat{z}&\coloneqq \frac{1}{L}\sum_{i=0}^{L-1}\hat{Z}_i^{(-1)^i}\notag\\
&=\frac{1}{L}(\hat{Z}_0+\hat{Z}_1^\dagger+\hat{Z}_2+\hat{Z}_3^\dagger+\dots+\hat{Z}_{L-2}^\dagger+\hat{Z}_{L-1}).\label{defOP2}
\end{align}
This nonzero expectation value is allowed by the absence of an exact spin-rotation symmetry.

Despite the lack of symmetry, we observe that the energy difference $\Delta_1\coloneqq E_1-E_0$ between the ground state and the first excited state vanishes at $g=1$, implying the presence a quantum phase transition of two phases at this point. See the inset of Fig.~\ref{fig3} (a). This is expected from our results for the even $L$ case, where a transition from an ordered phase ($1> g\geq0$) to a disordered phase ($g>1$) occurs at $g=1$. The thermodynamic behaviors, such as the presence or absence of a phase transition, must be insensitive to the detailed choice of the system size.

Indeed, even in this case, one can still form the Hamiltonian $\hat{H}^{(\ell_0=1)}(\epsilon)$ in Eq.~\eqref{He} with a symmetry breaking field field, where
$\hat{H}$ is given by Eq.~\eqref{generalizedmodel2} and $\hat{z}$ is given by Eq.~\eqref{defOP2}.
In Fig.~\ref{fig3} (c),(d), we show our numerical results on the expectation value $\mathrm{Re}[\langle\Phi_0^{(\ell_0=1)}(\epsilon)|\omega^*\hat{z}|\Phi_0^{(\ell_0=1)}(\epsilon)\rangle]$. The results for $1> g\geq0$ and $g>1$ clearly show qualitatively different behaviors. In particular, when $1> g\geq0$, we observe that $\text{Re}[\langle\Phi_0^{(1)}(\epsilon)|\omega^*\hat{z}|\Phi_0^{(1)}(\epsilon)\rangle]$ is negative for $0\leq\epsilon\ll \epsilon_*(L)$ and jumps to a positive value at $\epsilon=\epsilon_*(L)$.
The transition field $\epsilon_*(L)$ vanishes in the  large $L$ limit, implying that the large $L$ limit and the vanishing $\epsilon$ limit do not commute:
\begin{align}
&\lim_{\epsilon\rightarrow+0}
\lim_{L\rightarrow\infty}
\langle\Phi_0^{(\ell_0)}(\epsilon)|\hat{z}|\Phi_0^{(\ell_0)}(\epsilon)\rangle\notag\\
&\neq
\lim_{L\rightarrow\infty}
\lim_{\epsilon\rightarrow+0}
\langle\Phi_0^{(\ell_0)}(\epsilon)|\hat{z}|\Phi_0^{(\ell_0)}(\epsilon)\rangle.\label{noncommutative0}
\end{align}
Note that both hand sides are nonzero in this case, unlike Eq.~\eqref{noncommutative2}. 

The scaling of $\epsilon_*(L)$ is qualitatively different depending on the parity of $L$. As we saw in Sec.~\ref{secSSF1}, $\epsilon_*(L)$ decays exponentially with the system size in the standard model ($a=1$) and in the $a=N-1$ model with even $L$, while it only decays algebraically when the system size is odd for the $a=N-1$ case as shown  in Fig.~\ref{fig3} (e),(f) for $N=3$.  This behavior of $\epsilon_*(L)$ can also be explained by focusing on low-energy states.
Here we consider only two states, $|\Phi_0\rangle$ and $\hat{X}^{\ell_0}|\Phi_0\rangle$. At $\epsilon=0$, the energy expectation value of the latter state is greater than the former one by an amount $\Delta\sim O(1)$. However, the latter state is favored by the symmetry-breaking field. This suggests that the transition occurs at
\begin{align}
\epsilon_*(L)\simeq\frac{\Delta}{\big[1-\cos(\tfrac{2\pi \ell_0}{N})\big]Lm}\propto \frac{1}{L}\quad (\ell_0\neq0).
\end{align}

\begin{figure}[t]
\includegraphics[width=\columnwidth]{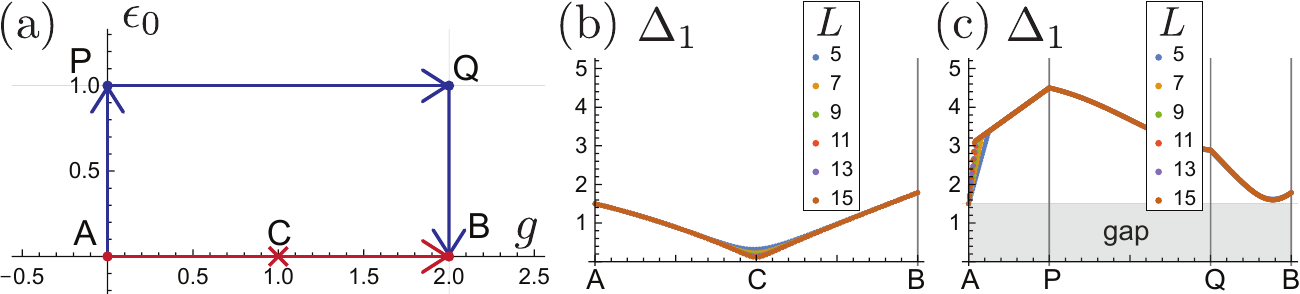}
\caption{\label{fig4} 
Exact-diagonalization results for the $(N,a)=(3,2)$ case.
(a): Paths in the $(g,\epsilon_0)$ plane. When A: $(0,0)$ is directly connected to B: $(2,0)$ (red arrow), a quantum phase transition occurs at  C: $(1,0)$.
However, when bypassed via P: $(0,1)$ and Q: $(1,1)$ (blue arrows), the transition can be avoided.
(b),(c): $\Delta_1=E_1-E_0$ for the red path [(b)] and the blue path [(c)] in the panel (a). To check the convergence as a function of the system size, results for $L=5,7,\cdots,15$ are shown.
}
\end{figure}

\subsubsection{Avoiding gap closing} 
When the system size $L$ is odd in the $a=N-1$ case, no exact symmetry of the model in Eq.~\eqref{generalizedmodel2} prohibits us from adding the longitudinal magnetic field term 
\begin{align}
\hat{H}(\epsilon_0)=\hat{H}-\frac{\epsilon_0}{2}\sum_{i=0}^{L-1}(\hat{Z}_i+\hat{Z}_i^\dagger).
\end{align}
When $\epsilon_0\neq0$, the unique ground state for $g=0$ and $g=2$ can be smoothly connected without a gap closing, as demonstrated numerically by Fig.~\ref{fig3} for $N=3$.

This observation suggests that the gap closing and the phase transition in odd system size were protected by the symmetry for the even system size. Namely, reference to the even $L$ system was mandatory for the discussion of odd $L$ system.

\subsection{$a=0\mod\mathrm{rad}(N)$}
As the last example, let us discuss the case in which the ground state for $g=0$ and $g=\infty$ can be adiabatically connected to each other without a gap closing, implying the uniqueness of the phase and the absence of any sort of phase transitions.

When $N$ is factorized as $N=\prod_{j}p_j^{r_j}$, the radical of $N$ is defined as $\mathrm{rad}(N)\coloneqq\prod_{j}p_j$. For example, $\mathrm{rad}(N)=2$ for $N=4=2^2$.  When $a$ is a multiple of $\mathrm{rad}(N)$, the $\mathbb{Z}_N$ symmetry is absent [i.e., $n$ in Eq.~\eqref{defn} is $N$] regardless of the system size $L$. As a consequence, the ground state is unique and excitations are gapped regardless of $g$.

\begin{figure}[t]
\includegraphics[width=\columnwidth]{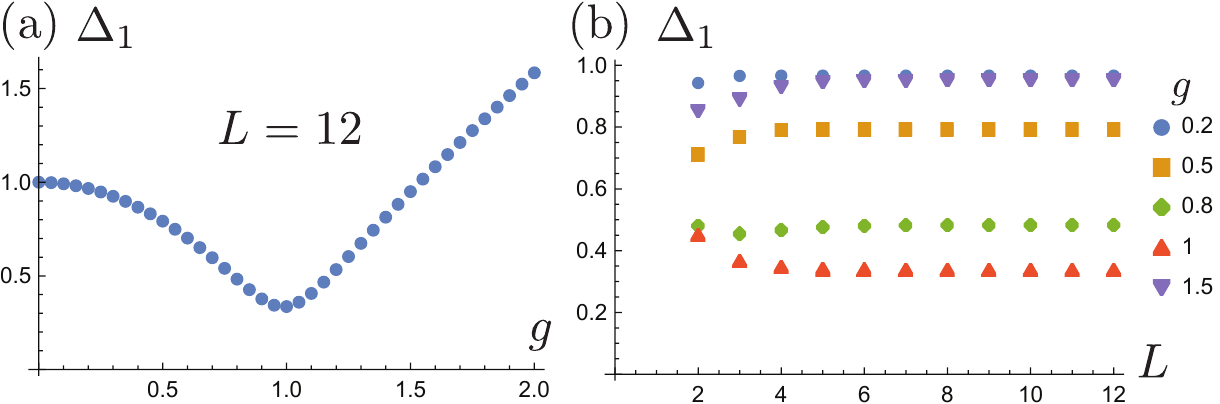}
\caption{\label{fig5}
Exact-diagonalization results for the $(N,a)=(4,2)$ case.
$\Delta_1=E_1-E_0$ as a function of (a) $g$ and (b) $L$.
}
\end{figure}

The simplest example of this situation is when $N=4$ and $a=2$. We show our numerical results in Fig.~\ref{fig5}. Clearly, the gap remains open as $g$ is changed from $0$ to $2$ for any value of $L$.

\section{Conclusion}
In this work, we introduced a generalized $N$-state clock model which contains an integer parameter $a=1,2,\cdots,N$.
The original $N$-state clock model corresponds to $a=1$.
When $a\neq1$,  the symmetry and the ground state degeneracy under the periodic boundary condition strongly depend on the system size $L$. 

In particular, when $a$ is $N-1$ and both $N$ and $L$ is odd, the spin-rotation symmetry is absent and the ground state is unique even in the ordered phase, despite the fact that a spontaneous ``symmetry" breaking is suggested by the non-commutativity of the large system size limit and the vanishing external field limit.  In contrast, when $L$ is even, the same model has the $\mathbb{Z}_N$ symmetry and exhibits spontaneous symmetry breaking in the standard manner.  Since thermodynamic properties should be insensitive to the details of the systems size or the boundary condition, this model with odd $L$ should be counted as an example of spontaneous symmetry breaking without exact symmetry or degeneracy. Indeed, we numerically found a gap closing at $g=1$.

The $a=N-1$ model can be mapped to the standard  $N$-state clock model ($a=+1$) by the $\hat{C}$ transformation in Eq.~\eqref{defC} for spins on even sites.
When $L$ is odd, this transformation introduces a defect for the spins at $i=L$ and $i=1$, which may be viewed as a boundary condition $\hat{Z}_{L}^\dagger=\hat{Z}_0$. This defect breaks the translation symmetry and the $\mathbb{Z}_N$ symmetry, explaining the absence of the ground state degeneracy.  However, other values of $a$ cannot be mapped to the $a=1$ model and the degeneracy pattern cannot be understood in this way. 

Although our model itself might be difficult to be realized in experiments, the importance of our example lies in the fact that it exemplifies the coexistence of a spontaneous symmetry breaking and a unique ground state with a finite excitation gap in a translationally invariant spin model with short-ranged interactions. Knowing this possibility is particularly important when one investigates interacting spin model numerically --- one often concludes the absence of any spontaneous symmetry breaking based on the uniqueness of the ground state, but our example draws cautions in such a reasoning.

\begin{acknowledgements}
We thank Yohei Fuji, Seishiro Ono, Hirokazu Kobayashi, Zijian Xiong, and Hosho Katsura for useful discussions.

The work of H.W. is supported by JSPS KAKENHI Grant No.~JP20H01825 and JP21H01789.
\end{acknowledgements}

\bibliography{ref.bib}

\begin{thebibliography}{22}%
\makeatletter
\providecommand \@ifxundefined [1]{%
 \@ifx{#1\undefined}
}%
\providecommand \@ifnum [1]{%
 \ifnum #1\expandafter \@firstoftwo
 \else \expandafter \@secondoftwo
 \fi
}%
\providecommand \@ifx [1]{%
 \ifx #1\expandafter \@firstoftwo
 \else \expandafter \@secondoftwo
 \fi
}%
\providecommand \natexlab [1]{#1}%
\providecommand \enquote  [1]{``#1''}%
\providecommand \bibnamefont  [1]{#1}%
\providecommand \bibfnamefont [1]{#1}%
\providecommand \citenamefont [1]{#1}%
\providecommand \href@noop [0]{\@secondoftwo}%
\providecommand \href [0]{\begingroup \@sanitize@url \@href}%
\providecommand \@href[1]{\@@startlink{#1}\@@href}%
\providecommand \@@href[1]{\endgroup#1\@@endlink}%
\providecommand \@sanitize@url [0]{\catcode `\\12\catcode `\$12\catcode
  `\&12\catcode `\#12\catcode `\^12\catcode `\_12\catcode `\%12\relax}%
\providecommand \@@startlink[1]{}%
\providecommand \@@endlink[0]{}%
\providecommand \url  [0]{\begingroup\@sanitize@url \@url }%
\providecommand \@url [1]{\endgroup\@href {#1}{\urlprefix }}%
\providecommand \urlprefix  [0]{URL }%
\providecommand \Eprint [0]{\href }%
\providecommand \doibase [0]{http://dx.doi.org/}%
\providecommand \selectlanguage [0]{\@gobble}%
\providecommand \bibinfo  [0]{\@secondoftwo}%
\providecommand \bibfield  [0]{\@secondoftwo}%
\providecommand \translation [1]{[#1]}%
\providecommand \BibitemOpen [0]{}%
\providecommand \bibitemStop [0]{}%
\providecommand \bibitemNoStop [0]{.\EOS\space}%
\providecommand \EOS [0]{\spacefactor3000\relax}%
\providecommand \BibitemShut  [1]{\csname bibitem#1\endcsname}%
\let\auto@bib@innerbib\@empty
\bibitem [{\citenamefont {Auerbach}(1998)}]{auerbach1998interacting}%
  \BibitemOpen
  \bibfield  {author} {\bibinfo {author} {\bibfnamefont {A.}~\bibnamefont
  {Auerbach}},\ }\href@noop {} {\emph {\bibinfo {title} {Interacting electrons
  and quantum magnetism}}}\ (\bibinfo  {publisher} {Springer},\ \bibinfo {year}
  {1998})\BibitemShut {NoStop}%
\bibitem [{\citenamefont {Altland}\ and\ \citenamefont
  {Simons}(2010)}]{altland_simons_2010}%
  \BibitemOpen
  \bibfield  {author} {\bibinfo {author} {\bibfnamefont {A.}~\bibnamefont
  {Altland}}\ and\ \bibinfo {author} {\bibfnamefont {B.~D.}\ \bibnamefont
  {Simons}},\ }\href {\doibase 10.1017/CBO9780511789984} {\emph {\bibinfo
  {title} {Condensed Matter Field Theory}}},\ \bibinfo {edition} {2nd}\ ed.\
  (\bibinfo  {publisher} {Cambridge University Press},\ \bibinfo {year}
  {2010})\BibitemShut {NoStop}%
\bibitem [{\citenamefont {Tasaki}(2020)}]{TasakiBook}%
  \BibitemOpen
  \bibfield  {author} {\bibinfo {author} {\bibfnamefont {H.}~\bibnamefont
  {Tasaki}},\ }\href@noop {} {\emph {\bibinfo {title} {Physics and mathematics
  of quantum many-body systems}}}\ (\bibinfo  {publisher} {Springer},\ \bibinfo
  {year} {2020})\BibitemShut {NoStop}%
\bibitem [{\citenamefont {Sachdev}(2011)}]{SachdevBook}%
  \BibitemOpen
  \bibfield  {author} {\bibinfo {author} {\bibfnamefont {S.}~\bibnamefont
  {Sachdev}},\ }\href@noop {} {\emph {\bibinfo {title} {Quantum Phase
  Transitions}}},\ \bibinfo {edition} {2nd}\ ed.\ (\bibinfo  {publisher}
  {Cambridge University Press},\ \bibinfo {year} {2011})\BibitemShut {NoStop}%
\bibitem [{\citenamefont {Bernien}\ \emph {et~al.}(2017)\citenamefont
  {Bernien}, \citenamefont {Schwartz}, \citenamefont {Keesling}, \citenamefont
  {Levine}, \citenamefont {Omran}, \citenamefont {Pichler}, \citenamefont
  {Choi}, \citenamefont {Zibrov}, \citenamefont {Endres}, \citenamefont
  {Greiner}, \citenamefont {Vuleti{\'c}},\ and\ \citenamefont
  {Lukin}}]{Bernien2017}%
  \BibitemOpen
  \bibfield  {author} {\bibinfo {author} {\bibfnamefont {H.}~\bibnamefont
  {Bernien}}, \bibinfo {author} {\bibfnamefont {S.}~\bibnamefont {Schwartz}},
  \bibinfo {author} {\bibfnamefont {A.}~\bibnamefont {Keesling}}, \bibinfo
  {author} {\bibfnamefont {H.}~\bibnamefont {Levine}}, \bibinfo {author}
  {\bibfnamefont {A.}~\bibnamefont {Omran}}, \bibinfo {author} {\bibfnamefont
  {H.}~\bibnamefont {Pichler}}, \bibinfo {author} {\bibfnamefont
  {S.}~\bibnamefont {Choi}}, \bibinfo {author} {\bibfnamefont {A.~S.}\
  \bibnamefont {Zibrov}}, \bibinfo {author} {\bibfnamefont {M.}~\bibnamefont
  {Endres}}, \bibinfo {author} {\bibfnamefont {M.}~\bibnamefont {Greiner}},
  \bibinfo {author} {\bibfnamefont {V.}~\bibnamefont {Vuleti{\'c}}}, \ and\
  \bibinfo {author} {\bibfnamefont {M.~D.}\ \bibnamefont {Lukin}},\ }\href
  {\doibase 10.1038/nature24622} {\bibfield  {journal} {\bibinfo  {journal}
  {Nature}\ }\textbf {\bibinfo {volume} {551}},\ \bibinfo {pages} {579}
  (\bibinfo {year} {2017})}\BibitemShut {NoStop}%
\bibitem [{\citenamefont {Whitsitt}\ \emph {et~al.}(2018)\citenamefont
  {Whitsitt}, \citenamefont {Samajdar},\ and\ \citenamefont
  {Sachdev}}]{PhysRevB.98.205118}%
  \BibitemOpen
  \bibfield  {author} {\bibinfo {author} {\bibfnamefont {S.}~\bibnamefont
  {Whitsitt}}, \bibinfo {author} {\bibfnamefont {R.}~\bibnamefont {Samajdar}},
  \ and\ \bibinfo {author} {\bibfnamefont {S.}~\bibnamefont {Sachdev}},\ }\href
  {\doibase 10.1103/PhysRevB.98.205118} {\bibfield  {journal} {\bibinfo
  {journal} {Phys. Rev. B}\ }\textbf {\bibinfo {volume} {98}},\ \bibinfo
  {pages} {205118} (\bibinfo {year} {2018})}\BibitemShut {NoStop}%
\bibitem [{\citenamefont {Watanabe}\ \emph {et~al.}(2022)\citenamefont
  {Watanabe}, \citenamefont {Cheng},\ and\ \citenamefont {Fuji}}]{2211.00299}%
  \BibitemOpen
  \bibfield  {author} {\bibinfo {author} {\bibfnamefont {H.}~\bibnamefont
  {Watanabe}}, \bibinfo {author} {\bibfnamefont {M.}~\bibnamefont {Cheng}}, \
  and\ \bibinfo {author} {\bibfnamefont {Y.}~\bibnamefont {Fuji}},\ }\href
  {https://arxiv.org/abs2211.00299} {\bibfield  {journal} {\bibinfo  {journal}
  {arXiv:2211.00299}\ } (\bibinfo {year} {2022})}\BibitemShut {NoStop}%
\bibitem [{\citenamefont {Dennis}\ \emph {et~al.}(2002)\citenamefont {Dennis},
  \citenamefont {Kitaev}, \citenamefont {Landahl},\ and\ \citenamefont
  {Preskill}}]{doi:10.1063/1.1499754}%
  \BibitemOpen
  \bibfield  {author} {\bibinfo {author} {\bibfnamefont {E.}~\bibnamefont
  {Dennis}}, \bibinfo {author} {\bibfnamefont {A.}~\bibnamefont {Kitaev}},
  \bibinfo {author} {\bibfnamefont {A.}~\bibnamefont {Landahl}}, \ and\
  \bibinfo {author} {\bibfnamefont {J.}~\bibnamefont {Preskill}},\ }\href
  {\doibase 10.1063/1.1499754} {\bibfield  {journal} {\bibinfo  {journal}
  {Journal of Mathematical Physics}\ }\textbf {\bibinfo {volume} {43}},\
  \bibinfo {pages} {4452} (\bibinfo {year} {2002})}\BibitemShut {NoStop}%
\bibitem [{\citenamefont {Kitaev}(2003)}]{KITAEV20032}%
  \BibitemOpen
  \bibfield  {author} {\bibinfo {author} {\bibfnamefont {A.}~\bibnamefont
  {Kitaev}},\ }\href {\doibase https://doi.org/10.1016/S0003-4916(02)00018-0}
  {\bibfield  {journal} {\bibinfo  {journal} {Annals of Physics}\ }\textbf
  {\bibinfo {volume} {303}},\ \bibinfo {pages} {2} (\bibinfo {year}
  {2003})}\BibitemShut {NoStop}%
\bibitem [{\citenamefont {Dong}\ \emph {et~al.}(2016)\citenamefont {Dong},
  \citenamefont {Li},\ and\ \citenamefont {Chen}}]{Dong_2016}%
  \BibitemOpen
  \bibfield  {author} {\bibinfo {author} {\bibfnamefont {J.-J.}\ \bibnamefont
  {Dong}}, \bibinfo {author} {\bibfnamefont {P.}~\bibnamefont {Li}}, \ and\
  \bibinfo {author} {\bibfnamefont {Q.-H.}\ \bibnamefont {Chen}},\ }\href
  {\doibase 10.1088/1742-5468/2016/11/113102} {\bibfield  {journal} {\bibinfo
  {journal} {Journal of Statistical Mechanics: Theory and Experiment}\ }\textbf
  {\bibinfo {volume} {2016}},\ \bibinfo {pages} {113102} (\bibinfo {year}
  {2016})}\BibitemShut {NoStop}%
\bibitem [{\citenamefont {Pelissetto}\ \emph {et~al.}(2018)\citenamefont
  {Pelissetto}, \citenamefont {Rossini},\ and\ \citenamefont
  {Vicari}}]{PhysRevE.98.032124}%
  \BibitemOpen
  \bibfield  {author} {\bibinfo {author} {\bibfnamefont {A.}~\bibnamefont
  {Pelissetto}}, \bibinfo {author} {\bibfnamefont {D.}~\bibnamefont {Rossini}},
  \ and\ \bibinfo {author} {\bibfnamefont {E.}~\bibnamefont {Vicari}},\ }\href
  {\doibase 10.1103/PhysRevE.98.032124} {\bibfield  {journal} {\bibinfo
  {journal} {Phys. Rev. E}\ }\textbf {\bibinfo {volume} {98}},\ \bibinfo
  {pages} {032124} (\bibinfo {year} {2018})}\BibitemShut {NoStop}%
\bibitem [{\citenamefont {Fishman}\ \emph {et~al.}(2022)\citenamefont
  {Fishman}, \citenamefont {White},\ and\ \citenamefont
  {Stoudenmire}}]{itensor}%
  \BibitemOpen
  \bibfield  {author} {\bibinfo {author} {\bibfnamefont {M.}~\bibnamefont
  {Fishman}}, \bibinfo {author} {\bibfnamefont {S.~R.}\ \bibnamefont {White}},
  \ and\ \bibinfo {author} {\bibfnamefont {E.~M.}\ \bibnamefont
  {Stoudenmire}},\ }\href {\doibase 10.21468/SciPostPhysCodeb.4} {\bibfield
  {journal} {\bibinfo  {journal} {SciPost Phys. Codebases}\ ,\ \bibinfo {pages}
  {4}} (\bibinfo {year} {2022})}\BibitemShut {NoStop}%
\bibitem [{\citenamefont {Ortiz}\ \emph {et~al.}(2012)\citenamefont {Ortiz},
  \citenamefont {Cobanera},\ and\ \citenamefont {Nussinov}}]{ORTIZ2012780}%
  \BibitemOpen
  \bibfield  {author} {\bibinfo {author} {\bibfnamefont {G.}~\bibnamefont
  {Ortiz}}, \bibinfo {author} {\bibfnamefont {E.}~\bibnamefont {Cobanera}}, \
  and\ \bibinfo {author} {\bibfnamefont {Z.}~\bibnamefont {Nussinov}},\ }\href
  {\doibase https://doi.org/10.1016/j.nuclphysb.2011.09.012} {\bibfield
  {journal} {\bibinfo  {journal} {Nuclear Physics B}\ }\textbf {\bibinfo
  {volume} {854}},\ \bibinfo {pages} {780} (\bibinfo {year}
  {2012})}\BibitemShut {NoStop}%
\bibitem [{\citenamefont {Moran}\ \emph {et~al.}(2017)\citenamefont {Moran},
  \citenamefont {Pellegrino}, \citenamefont {Slingerland},\ and\ \citenamefont
  {Kells}}]{PhysRevB.95.235127}%
  \BibitemOpen
  \bibfield  {author} {\bibinfo {author} {\bibfnamefont {N.}~\bibnamefont
  {Moran}}, \bibinfo {author} {\bibfnamefont {D.}~\bibnamefont {Pellegrino}},
  \bibinfo {author} {\bibfnamefont {J.~K.}\ \bibnamefont {Slingerland}}, \ and\
  \bibinfo {author} {\bibfnamefont {G.}~\bibnamefont {Kells}},\ }\href
  {\doibase 10.1103/PhysRevB.95.235127} {\bibfield  {journal} {\bibinfo
  {journal} {Phys. Rev. B}\ }\textbf {\bibinfo {volume} {95}},\ \bibinfo
  {pages} {235127} (\bibinfo {year} {2017})}\BibitemShut {NoStop}%
\bibitem [{\citenamefont {Sun}\ \emph {et~al.}(2019)\citenamefont {Sun},
  \citenamefont {Vekua}, \citenamefont {Cobanera},\ and\ \citenamefont
  {Ortiz}}]{PhysRevB.100.094428}%
  \BibitemOpen
  \bibfield  {author} {\bibinfo {author} {\bibfnamefont {G.}~\bibnamefont
  {Sun}}, \bibinfo {author} {\bibfnamefont {T.}~\bibnamefont {Vekua}}, \bibinfo
  {author} {\bibfnamefont {E.}~\bibnamefont {Cobanera}}, \ and\ \bibinfo
  {author} {\bibfnamefont {G.}~\bibnamefont {Ortiz}},\ }\href {\doibase
  10.1103/PhysRevB.100.094428} {\bibfield  {journal} {\bibinfo  {journal}
  {Phys. Rev. B}\ }\textbf {\bibinfo {volume} {100}},\ \bibinfo {pages}
  {094428} (\bibinfo {year} {2019})}\BibitemShut {NoStop}%
\bibitem [{\citenamefont {Pfeuty}(1970)}]{PFEUTY197079}%
  \BibitemOpen
  \bibfield  {author} {\bibinfo {author} {\bibfnamefont {P.}~\bibnamefont
  {Pfeuty}},\ }\href {\doibase https://doi.org/10.1016/0003-4916(70)90270-8}
  {\bibfield  {journal} {\bibinfo  {journal} {Annals of Physics}\ }\textbf
  {\bibinfo {volume} {57}},\ \bibinfo {pages} {79} (\bibinfo {year}
  {1970})}\BibitemShut {NoStop}%
\bibitem [{\citenamefont {A.~Kaplan}\ \emph {et~al.}(1989)\citenamefont
  {A.~Kaplan}, \citenamefont {Horsch},\ and\ \citenamefont {von~der
  Linden}}]{doi:10.1143/JPSJ.58.3894}%
  \BibitemOpen
  \bibfield  {author} {\bibinfo {author} {\bibfnamefont {T.}~\bibnamefont
  {A.~Kaplan}}, \bibinfo {author} {\bibfnamefont {P.}~\bibnamefont {Horsch}}, \
  and\ \bibinfo {author} {\bibfnamefont {W.}~\bibnamefont {von~der Linden}},\
  }\href {\doibase 10.1143/JPSJ.58.3894} {\bibfield  {journal} {\bibinfo
  {journal} {Journal of the Physical Society of Japan}\ }\textbf {\bibinfo
  {volume} {58}},\ \bibinfo {pages} {3894} (\bibinfo {year}
  {1989})}\BibitemShut {NoStop}%
\bibitem [{\citenamefont {Cabrera}\ and\ \citenamefont
  {Jullien}(1987)}]{PhysRevB.35.7062}%
  \BibitemOpen
  \bibfield  {author} {\bibinfo {author} {\bibfnamefont {G.~G.}\ \bibnamefont
  {Cabrera}}\ and\ \bibinfo {author} {\bibfnamefont {R.}~\bibnamefont
  {Jullien}},\ }\href {\doibase 10.1103/PhysRevB.35.7062} {\bibfield  {journal}
  {\bibinfo  {journal} {Phys. Rev. B}\ }\textbf {\bibinfo {volume} {35}},\
  \bibinfo {pages} {7062} (\bibinfo {year} {1987})}\BibitemShut {NoStop}%
\bibitem [{\citenamefont {Lieb}\ \emph {et~al.}(1961)\citenamefont {Lieb},
  \citenamefont {Schultz},\ and\ \citenamefont {Mattis}}]{LiebSchultzMattis}%
  \BibitemOpen
  \bibfield  {author} {\bibinfo {author} {\bibfnamefont {E.}~\bibnamefont
  {Lieb}}, \bibinfo {author} {\bibfnamefont {T.}~\bibnamefont {Schultz}}, \
  and\ \bibinfo {author} {\bibfnamefont {D.}~\bibnamefont {Mattis}},\ }\href
  {\doibase https://doi.org/10.1016/0003-4916(61)90115-4} {\bibfield  {journal}
  {\bibinfo  {journal} {Annals of Physics}\ }\textbf {\bibinfo {volume} {16}},\
  \bibinfo {pages} {407} (\bibinfo {year} {1961})}\BibitemShut {NoStop}%
\bibitem [{\citenamefont {McCoy}(2009)}]{McCoyBook}%
  \BibitemOpen
  \bibfield  {author} {\bibinfo {author} {\bibfnamefont {B.~M.}\ \bibnamefont
  {McCoy}},\ }\href {\doibase 10.1093/acprof:oso/9780199556632.001.0001} {\emph
  {\bibinfo {title} {{Advanced Statistical Mechanics}}}}\ (\bibinfo
  {publisher} {Oxford University Press},\ \bibinfo {year} {2009})\BibitemShut
  {NoStop}%
\bibitem [{\citenamefont {Suzuki}\ \emph {et~al.}(2012)\citenamefont {Suzuki},
  \citenamefont {Inoue},\ and\ \citenamefont {Chakrabarti}}]{SuzukiBook}%
  \BibitemOpen
  \bibfield  {author} {\bibinfo {author} {\bibfnamefont {S.}~\bibnamefont
  {Suzuki}}, \bibinfo {author} {\bibfnamefont {J.-i.}\ \bibnamefont {Inoue}}, \
  and\ \bibinfo {author} {\bibfnamefont {B.~K.}\ \bibnamefont {Chakrabarti}},\
  }\href@noop {} {\emph {\bibinfo {title} {Quantum Ising phases and transitions
  in transverse Ising models}}},\ Vol.\ \bibinfo {volume} {862}\ (\bibinfo
  {publisher} {Springer},\ \bibinfo {year} {2012})\BibitemShut {NoStop}%
\bibitem [{\citenamefont {Barber}\ and\ \citenamefont
  {Fisher}(1973)}]{BARBER19731}%
  \BibitemOpen
  \bibfield  {author} {\bibinfo {author} {\bibfnamefont {M.~N.}\ \bibnamefont
  {Barber}}\ and\ \bibinfo {author} {\bibfnamefont {M.~E.}\ \bibnamefont
  {Fisher}},\ }\href {\doibase https://doi.org/10.1016/0003-4916(73)90409-0}
  {\bibfield  {journal} {\bibinfo  {journal} {Annals of Physics}\ }\textbf
  {\bibinfo {volume} {77}},\ \bibinfo {pages} {1} (\bibinfo {year}
  {1973})}\BibitemShut {NoStop}%
\end{thebibliography}%
\clearpage
\appendix
\section{Exact solution for $N=2$}
\label{appJW}
We review the exact solution of the transverse-field Ising model via  the Jordan-Wigner transformation following Refs.~\onlinecite{LiebSchultzMattis,PFEUTY197079,PhysRevB.35.7062,McCoyBook,SuzukiBook}.
\subsection{Jordan-Wigner transformation}
$N=2$ level spins can be represented by fermion operators:
\begin{align}
&\hat{X}_i=(\hat{f}_i+\hat{f}_i^\dagger) \prod_{i'=0}^{i-1}(-1)^{\hat{f}_{i'}^\dagger \hat{f}_{i'}},\label{JW1}\\
&\hat{Z}_i=(-1)^{\hat{f}_i^\dagger \hat{f}_i}=1-2\hat{f}_i^\dagger \hat{f}_i=\hat{f}_i\hat{f}_i^\dagger-\hat{f}_i^\dagger \hat{f}_i,\label{JW3}
\end{align}
where $\prod_{i'=0}^{i-1}(-1)^{\hat{f}_{i'}^\dagger \hat{f}_{i'}}=1$ when $i=0$. 
The product state $|\Phi_0\rangle=\bigotimes_{i=0}^{L-1}|{1}\rangle_i$ is mapped to the Fock vacuum $|0\rangle$.
The definition of the Jordan-Wigner transformation here is slightly different from the standard one~\cite{LiebSchultzMattis,PFEUTY197079,McCoyBook,SuzukiBook} but we find that this choice is more useful in that expressions in Eqs.~\eqref{P0} and \eqref{P1} below do not depend on the parity of $L$.

Interchanging the role of $\hat{X}_i$ and $\hat{Z}_i$, we find
\begin{align}
\hat{H}&=-\sum_{i=0}^{L-1}\hat{X}_{i+1}\hat{X}_i-g\sum_{i=0}^{L-1}\hat{Z}_i\notag\\
&=-\sum_{i=0}^{L-2}(\hat{f}_{i+1}^\dagger+\hat{f}_{i+1})(\hat{f}_i-\hat{f}_i^\dagger)\notag\\
&\quad-[-(-1)^{\hat{N}}](\hat{f}_{0}^\dagger+\hat{f}_{0})(\hat{f}_{L-1}-\hat{f}_{L-1}^\dagger)\notag\\
&\quad-g\sum_{i=0}^{L-1}(1-2\hat{f}_i^\dagger \hat{f}_i).
\end{align}
In the $(-1)^{\hat{N}}=+1$ ($-1$) sector, we set the boundary condition to be anti-periodic $\hat{f}_{L+i}=-\hat{f}_{i}$ (periodic $\hat{f}_{L+i}=\hat{f}_{i}$). With this understanding, the Hamiltonian can be rewritten as
\begin{align}
\hat{H}&=-\sum_{i=0}^{L-1}(\hat{f}_{i+1}^\dagger\hat{f}_i+\hat{f}_i^\dagger\hat{f}_{i+1}+\hat{f}_{i+1}\hat{f}_i+\hat{f}_i^\dagger\hat{f}_{i+1}^\dagger)\notag\\
&\quad+g\sum_{i=0}^{L-1}(\hat{f}_i^\dagger \hat{f}_i-\hat{f}_i\hat{f}_i^\dagger).
\end{align}
Introducing the Fourier transformation $\hat{f}_j^\dagger=L^{-1/2}\sum_{k}\hat{f}_k^\dagger e^{-ikj}$, where $k\in K_{\text{AP}}=\{(2j+1)\pi/L\}_{j=0}^{L-1}$ for the anti-periodic case and $k\in K_{\text{P}}=\{(2j)\pi/L\}_{j=0}^{L-1}$ for the periodic case, we find
\begin{align}
\hat{H}
&=\sum_k
\begin{pmatrix}
\hat{f}_k^\dagger &\hat{f}_{-k}
\end{pmatrix}
\begin{pmatrix}
g-\cos k & -i\sin k\\
i\sin k& -g+\cos k
\end{pmatrix}
\begin{pmatrix}
\hat{f}_k\\
\hat{f}_{-k}^\dagger
\end{pmatrix}\notag\\
&=\sum_k\varepsilon(k)
\begin{pmatrix}
\hat{f}_k^\dagger &\hat{f}_{-k}
\end{pmatrix}
(\cos2\phi_k\sigma_3+\sin2\phi_k\sigma_2)
\begin{pmatrix}
\hat{f}_k\\
\hat{f}_{-k}^\dagger
\end{pmatrix}\notag\\
&=\sum_k\varepsilon(k)
\begin{pmatrix}
\hat{\gamma}_k^\dagger &\hat{\gamma}_{-k}
\end{pmatrix}
\sigma_3\begin{pmatrix}
\hat{\gamma}_k\\
\hat{\gamma}_{-k}^\dagger
\end{pmatrix}\notag\\
&=\sum_k2\varepsilon_k\hat{\gamma}_k^\dagger \hat{\gamma}_k-\sum_k\varepsilon_k.\label{BDGdiag}
\end{align}
In the derivation, we defined 
\begin{align}
&\varepsilon(k)=\sqrt{(g-\cos k)^2+\sin^2k}=\sqrt{1+g^2-2g\cos k},\\
&\cos2\phi_k=\frac{g-\cos k}{\varepsilon_k},\quad\sin2\phi_k=\frac{\sin k}{\varepsilon_k},\\
&\begin{pmatrix}
\hat{\gamma}_k\\
\hat{\gamma}_{-k}^\dagger
\end{pmatrix}=
e^{-i\phi_k\sigma_1}
\begin{pmatrix}
\hat{f}_k\\
\hat{f}_{-k}^\dagger
\end{pmatrix}
=\begin{pmatrix}
\cos \phi_k & -i\sin \phi_k\\
-i\sin \phi_k& \cos \phi_k
\end{pmatrix}
\begin{pmatrix}
\hat{f}_k\\
\hat{f}_{-k}^\dagger
\end{pmatrix}.
\end{align}
The last expression of Eq.~\eqref{BDGdiag} implies that the ground states are those annihilated by $\hat{\gamma}_k$ for all $k$.

For $k=\pi$, $g-\cos k=g+1>0$ regardless of $g\geq0$. Hence, $\phi_{k=\pi}=0$ and $\hat{\gamma}_{k=\pi}=\hat{f}_{k=\pi}$.
On the other hand, for $k=0$, $g-\cos k=g-1>0$  and $\phi_{k=0}=0$ and $\hat{\gamma}_{k=0}=\hat{f}_{k=0}$ only when $g>1$. In contrast, when $1>g\geq0$, $\phi_{k=0}=\pi/2$ and $\hat{\gamma}_{k=0}=\hat{f}_{k=0}^\dagger$. 

\subsection{Ground state energy}
The state
\begin{align}
|\Phi_0\rangle=\prod_{k\in K_{\text{AP}}}(\cos \phi_k+i\sin \phi_k\hat{f}_k^\dagger\hat{f}_{-k}^\dagger)|0\rangle\label{P0}
\end{align}
satisfies $\hat{\gamma}_k|\Phi_0\rangle=0$ for any $k\in K_{\text{AP}}$ and hence is the ground state in the the even fermion parity sector. The energy eigenvalue is given by
\begin{align}
E_0(g)&=-\sum_{k\in K_{\text{AP}}}\varepsilon_k=-\sum_{j=0}^{L-1}\varepsilon(\tfrac{(2j+1)\pi}{L}).
\end{align}
This result is valid for any $g\geq0$.

On the other hand, the state
\begin{align}
|\Phi_1\rangle=\hat{f}_{k=0}^\dagger\prod_{k\in K_{\text{P}},k\neq0}(\cos \phi_k+i\sin \phi_k\hat{f}_k^\dagger\hat{f}_{-k}^\dagger)|0\rangle\label{P1}
\end{align}
is the ground state in the odd fermion parity sector.  It satisfies $\hat{\gamma}_k|\Phi_1\rangle=0$ for any $k\in K_{\text{P}}$  when $1>g\geq0$. When $1>g$, $\gamma_{k=0}|\Phi_1\rangle$ does not vanish but this state remains the ground state in this sector because $\varepsilon_k$ is the monotonically increasing function of $|k|$ in the range $0\leq |k|\leq \pi$. The energy eigenvalue is given by
\begin{align}
E_1(g)&=-\sum_{k\in K_{\text{P}}}\varepsilon_k=-\sum_{j=0}^{L-1}\varepsilon(\tfrac{2\pi j}{L}).
\end{align}
When $1>g$, $2(g-1)$ should be added to $E_1(g)$.

Comparing $E_0(g)$ and $E_1(g)$, we find $E_1(g)>E_0(g)$ whenever $g>0$. Thus $|\Phi_0\rangle$ is the ground state in the finite system and $|\Phi_1\rangle$ is the quasi-degenerate first excited state in a finite system. To evaluate the difference $E_1(g)-E_0(g)$ for $1>g\geq0$, we follow the prescription given in Ref.~\onlinecite{BARBER19731}. We first perform Fourier transformation:
\begin{align}
&\varepsilon_n=\int_{0}^{2\pi}\frac{dk}{2\pi}\varepsilon(k)e^{-ink}=\int_{0}^{2\pi}\frac{dk}{2\pi}\varepsilon(k)\cos nk,\\
&\varepsilon(k)=\sum_{n=-\infty}^\infty \varepsilon_ne^{ink}=\varepsilon_0+2\sum_{n=1}^\infty \varepsilon_n\cos nk.
\end{align}
In terms of these Fourier components, $E_0(g)$ and $E_1(g)$ can be expressed as
\begin{align}
E_0(g)&=-\sum_{n=-\infty}^\infty\varepsilon_ne^{i\tfrac{n\pi}{L}} \sum_{j=0}^{L-1}e^{i\tfrac{2n\pi }{L}j}=-L\sum_{n=-\infty}^\infty \varepsilon_{mL}(-1)^m\notag\\
&=-L\varepsilon_0-2L\sum_{m=1}^\infty \varepsilon_{mL}(-1)^m
\end{align}
and
\begin{align}
E_1(g)&=-\sum_{n=-\infty}^\infty \varepsilon_n \sum_{j=0}^{L-1}e^{i\tfrac{2\pi n}{L}j}=-L\sum_{m=-\infty}^\infty \varepsilon_{mL}\notag\\
&=-L\varepsilon_0-2L\sum_{m=1}^\infty \varepsilon_{mL}.
\end{align}

Therefore, we get
\begin{align}
&E_1(g)-E_0(g)=-4L\sum_{m=0}^\infty \varepsilon_{(1+2m)L}\notag\\
&=-4L\sum_{m=0}^\infty\int_{0}^{2\pi}\frac{dk}{2\pi}\varepsilon(k)e^{i(1+2m)Lk}.
\end{align}
To examine the asymptotic behavior of this integral, we introduce $\lambda>0$ by $\lambda=-\log g$ ($e^{-\lambda}=g$) so that
\begin{align}
e^{i(k+i\lambda)}=e^{ik}g,\quad e^{- i(k+i\lambda)}=e^{-ik}/g.
\end{align}
It follows that 
\begin{align}
&\varepsilon(k+i\lambda)=\sqrt{(e^{ik}-1)(e^{-ik}-g^2)},\\
&e^{i(1+2m)L(k+i\lambda)}=e^{i(1+2m)Lk}g^{(1+2m)L}.
\end{align}
Since $\varepsilon(z)e^{i(1+2m)Lz}$ is analytic when $|\mathrm{Im}z|\leq\lambda$, the integration path can be shifted from the real axis to $k+i\lambda$ with $k\in[0,2\pi]$. We find
\begin{align}
E_1(g)-E_0(g)=4L\sum_{m=0}^\infty g^{(1+2m)L}I_{(1+2m)L}(g),\label{Delta2}
\end{align}
where
\begin{align}
&I_L(g)\coloneqq-\int_{0}^{2\pi}\frac{dk}{2\pi}\sqrt{(e^{ik}-1)(e^{-ik}-g^2)}e^{iLk}\notag\\
&=\frac{\Gamma(L-\frac{1}{2})}{\sqrt{4\pi}\Gamma(L+1)}{}_2F_1(-\tfrac{1}{2},L-\tfrac{1}{2};L+1;g^2)
\end{align}
and ${}_2F_1(a,b;c;z)$ is the hypergeometric function.  The sum over $m$ in Eq.~\eqref{Delta2} is clearly dominated by the $m=0$ contribution.
Therefore, we obtain the asymptotic form for $L\rightarrow\infty$:
\begin{align}
&E_1(g)-E_0(g)\notag\\
&\simeq\frac{2L\,\Gamma(L-\frac{1}{2})g^L}{\sqrt{\pi}\Gamma(L+1)}{}_2F_1(-\tfrac{1}{2},L-\tfrac{1}{2};L+1;g^2)+O(g^{3L})\notag\\
&\simeq\Big(1+\frac{3(1+g^2)}{8(1-g^2)L}+\frac{5(5g^4-22g^2+5)}{128(1-g^2)^2L^2}\notag\\
&\quad\quad+\frac{105(1+g^2)^3}{1024(1-g^2)^3L^3}+O(L^{-4})\Big) 2\sqrt{\frac{1-g^2}{\pi L}}g^L.
\end{align}
The leading term was previously derived in Ref.~\onlinecite{PhysRevB.35.7062}. The $O(L^{-m})$ ($m=1,2,3$) corrections were not found in literature, and higher order corrections can be computed in the same way.

\subsection{Long range correlation}
Next, let us investigate the correlation functions.
\begin{align}
&\hat{X}_i\hat{X}_{i+n}=(\hat{f}_i+\hat{f}_i^\dagger)\prod_{j=i}^{i+n-1}(-1)^{\hat{f}_{j}^\dagger \hat{f}_{j}}(\hat{f}_{i+n}+\hat{f}_{i+n}^\dagger)\notag\\
&=(\hat{f}_i^\dagger-\hat{f}_i)\prod_{j=i+1}^{i+n-1}(\hat{f}_{j}^\dagger+\hat{f}_{j})(\hat{f}_{j}^\dagger-\hat{f}_{j})(\hat{f}_{i+n}^\dagger+\hat{f}_{i+n})\notag\\
&=\hat{B}_i\hat{A}_{i+1}\hat{B}_{i+1}\hat{A}_{i+2}\cdots\hat{B}_{i+n-1}\hat{A}_{i+n},
\end{align}
where $\hat{A}_i\coloneqq\hat{f}_i^\dagger+\hat{f}_i$ and $\hat{B}_i\coloneqq\hat{f}_i^\dagger-\hat{f}_i$, which satisfy $\{\hat{A}_i,\hat{B}_j\}=0$ and $\{\hat{A}_i,\hat{A}_j\}=-\{\hat{B}_i,\hat{B}_j\}=2\delta_{ij}$. 

We are interested in the correlation function with respect to the ground state $|\Phi_0\rangle$. According to Wick's theorem, the correlation function can be decomposed into the sum of two point functions: 
\begin{align}
\langle\hat{X}_i\hat{X}_{i+n}\rangle&=\langle\hat{B}_i\hat{A}_{i+1}\hat{B}_{i+1}\hat{A}_{i+2}\cdots\hat{B}_{i+n-1}\hat{A}_{i+n}\rangle\notag\\
&=\sum_\sigma \mathrm{sgn}(\sigma)\prod_{j=i}^{i+n-1}\langle\hat{B}_j\hat{A}_{\sigma(j)+1}\rangle.\label{corre1}
\end{align}
Using
\begin{align}
&\hat{f}_i=\frac{1}{\sqrt{L}}\sum_{k\in K_{\text{AP}}}e^{iki}\left(\cos \phi_k\hat{\gamma}_k + i\sin \phi_k\hat{\gamma}_{-k}^\dagger\right)
\end{align}
and $\hat{\gamma}_k|\Phi_0\rangle=0$, we get
\begin{align}
&\langle\hat{f}_i^\dagger\hat{f}_j^\dagger\rangle=-\langle\hat{f}_i\hat{f}_j\rangle=\frac{i}{2L}\sum_{k\in K_{\text{AP}}}e^{ik(i-j)}\frac{\sin k}{\varepsilon_k},\\
&\langle\hat{f}_i\hat{f}_j^\dagger\rangle=\langle\hat{f}_j\hat{f}_i^\dagger\rangle=\frac{1}{2L}\sum_{k\in K_{\text{AP}}}e^{ik(i-j)}(1+\frac{g-\cos k}{\varepsilon_k}),\\
&\langle\hat{f}_i^\dagger\hat{f}_j\rangle=\langle\hat{f}_j^\dagger\hat{f}_i\rangle=\frac{1}{2L}\sum_{k\in K_{\text{AP}}}e^{ik(i-j)}(1-\frac{g-\cos k}{\varepsilon_k}),
\end{align}
from which we find
\begin{align}
\langle\hat{A}_i\hat{A}_j\rangle&=\langle\hat{f}_i^\dagger\hat{f}_j^\dagger+\hat{f}_i\hat{f}_j+\hat{f}_i^\dagger\hat{f}_j+\hat{f}_i\hat{f}_j^\dagger\rangle=\delta_{ij},\\
\langle\hat{B}_i\hat{B}_j\rangle&=\langle\hat{f}_i^\dagger\hat{f}_j^\dagger+\hat{f}_i\hat{f}_j-\hat{f}_i^\dagger\hat{f}_j-\hat{f}_i\hat{f}_j^\dagger\rangle=-\delta_{ij},\\
\langle\hat{B}_i\hat{A}_j\rangle&=\langle\hat{f}_i^\dagger\hat{f}_j^\dagger-\hat{f}_i\hat{f}_j+\hat{f}_i^\dagger\hat{f}_j-\hat{f}_i\hat{f}_j^\dagger\rangle=G_{j-i-1}.
\end{align}
Here we defined
\begin{align}
G_{m}&\coloneqq\langle\hat{B}_i\hat{A}_{m+i+1}\rangle=\frac{1}{L}\sum_{k\in K_{\text{AP}}}\frac{1-ge^{ik}}{\varepsilon_k}e^{ikm}\notag\\
&=\frac{1}{L}\sum_{k\in K_{\text{AP}}}\sqrt{\frac{1-ge^{ik}}{1-ge^{-ik}}}e^{ikm}\simeq \int_0^{2\pi}\frac{dk}{2\pi}G(e^{-ik})e^{ikm}
\end{align}
and
\begin{align}
G(e^{-ik})=\sqrt{\frac{1-ge^{ik}}{1-ge^{-ik}}}.
\end{align}
Therefore, the correlation function in Eq.~\eqref{corre1} can be written as the determinant of a Toeplitz matrix:
\begin{align}
&\langle\hat{X}_i\hat{X}_{i+n}\rangle\simeq\begin{vmatrix}
G_{0} & G_{-1} & G_{-2} & \cdots  & G_{1-n}\\
G_{1} & G_{0} & G_{-1} & \cdots & G_{2-n}\\
G_{2} & G_{1} & G_{0} &  \cdots & G_{3-n}\\
\vdots & \vdots & \vdots & & \vdots\\
G_{n-1} & G_{n-2} & G_{n-3} & \cdots & G_{0}\\
\end{vmatrix},
\end{align}
whose asymptoric behavior is given by the strong Szeg\"o limit theorem~\cite{McCoyBook,SuzukiBook}.
\begin{align}
\langle\hat{X}_i\hat{X}_{i+n}\rangle\simeq (e^{d_0})^n e^{\sum_{n=1}^\infty n d_n d_{-n}}=(1-g^2)^{1/4},
\end{align}
where
\begin{align}
&d_0=\int_0^{2\pi}\frac{dk}{2\pi}\log[G(e^{-ik})]=0,\\
&d_n=\int_0^{2\pi}\frac{dk}{2\pi}\log[G(e^{-ik})]e^{ikn}=\frac{g^{|n|}}{2n}.
\end{align}
This result implies~\cite{PFEUTY197079}
\begin{align}
\lim_{L\rightarrow\infty}m(g)=(1-g^2)^{1/8}.
\end{align}

\end{document}